\documentstyle[12pt,aasms4]{article}
\def\ea{{\it et al. }}
\def\etal{{\it et al.}}
\def\ltsima{$\; \buildrel < \over \sim \;$}
\def\simlt{\lower.5ex\hbox{\ltsima}}      
\def\gtsima{$\; \buildrel > \over \sim \;$}
\def\simgt{\lower.5ex\hbox{\gtsima}}       
\def\deg{$^\circ$}
\received{~~}
\accepted{~~}
\journalid{}{}
\articleid{}{}

\begin{document}

\title{Below the Lyman Edge: UV Polarimetry of Quasars}
\author{Anuradha Koratkar}
\affil{Space Telescope Science Institute, 3700 San Martin Dr., Baltimore,
 MD 21218}
\authoremail{koratkar@stsci.edu}
\author{Robert Antonucci}
\affil{Physics Department, University of California, Santa Barbara, CA93106}
\authoremail{ski@ginger.physics.ucsb.edu}
\author{Robert Goodrich}
\affil{Keck Observatory}
\authoremail{goodrich@keck.hawaii.edu}
\author{Alex Storrs}
\affil{Space Telescope Science Institute, 3700 San Martin Dr., Baltimore,
 MD 21218}
\authoremail{storrs@stsci.edu}

\begin{abstract}

The Lyman edge at 912 {\AA} is an important diagnostic region for
studying quasi-stellar objects (QSOs).  In particular, it reveals a
great deal about the physical conditions within the atmospheres of
accretion disks, a ubiquitous component of QSO theories.  A robust
prediction of accretion disk models is a significant polarization due
to electron scattering just longward (in wavelength) of the Lyman edge
because of the wavelength dependence of the Hydrogen absorption
opacity.  Observations of the Lyman edge regions of QSOs have shown
scant evidence for the predicted features---few QSOs show the broad,
partial Lyman edges expected to be common according to most theories,
and none show the high polarizations expected longward of the Lyman
edge.  Still, polarization spectra of a small number of QSOs have shown
a rising polarization (up to 20\%) at wavelengths {\it shortward} of
the Lyman edge. We have now doubled our sample of intermediate-redshift
QSOs observed with the {\it HST/FOS} spectropolarimeter to determine
the amount of polarization on both sides of the Lyman limit. For this
new sample of six objects, polarizations are low and mostly consistent
with zero below the Lyman edge.

Another important result of the new data is that it strengthens the
conclusion that quasars are generally not polarized significantly just
longward of the Lyman edge at $\sim$1000\AA\/.  There is no significant
statistical wavelength dependence to the polarization longward of the
Lyman edge indicating that simple plane-parallel atmospheres with
scattering-dominated opacity are not significant sources of UV flux in
quasars.

\end{abstract}

\keywords{accretion, accretion disks -- polarization -- quasars:general --
ultraviolet:galaxies}

\section{Introduction}

One of the fundamental components of most theories of quasi-stellar
objects (QSOs) is an accretion disk.  As gas is fed into the central
regions of the QSO, residual angular momentum causes the gas to
naturally settle into a disk.  While most theories predict the
formation of such a disk, few address the observational consequences of
the disk models in detail.  These theoretical studies have found that
the Lyman edge at 912 {\AA} is a powerful diagnostic feature for the
physical characteristics of the disk.

The simplest disk models (quasi-static with viscous dissipation at
large optical depth) predict Lyman edges in either emission or
absorption, depending on the viewing angle and the physical details of
the disk atmosphere.  Such Lyman edges would be broadened by rotation
of the disk and by general relativistic effects as light passes close
to the central black hole.  In most QSOs such edges are not seen,
although Koratkar, Kinney, \& Bohlin (1992) found a small number of
candidate ``partial edges'' in IUE data.

A second disk signature is the linear polarization, $P$, of the
continuum from the disk.  A purely scattering atmosphere will produce
high polarization perpendicular to the disk axis.  Again, this
signature is not seen in any QSOs; in fact generally QSOs show {\it
low} optical polarization {\it parallel} to the inferred disk axis.
Laor, Netzer, \& Piran (1990) attempted to show that a disk atmosphere
should have significant absorptive opacity, and thus can produce
dramatically lower optical polarization (albeit still perpendicular to
the disk axis and thus inconsistent with the observations).  A more
robust prediction according to their work, however, is a rise in
polarization with decreasing wavelength from the optical into the UV.
Just longward of the Lyman edge $P$ is highest, since it is at these
wavelengths that scattering best competes with absorption.  Just
shortward of the Lyman edge, as the absorption opacity increases, $P$
should drop again.  According to Laor \etal, this polarization
signature should appear even when no disk signature is seen in total
flux.

In our first polarization study of three of the rare objects known from
IUE spectra to have partial Lyman edges at the systemic redshifts, we
found low polarizations longward of the edges, so we did not confirm
the Laor \ea prediction.  We surprisingly did find high polarization
{\it shortward} of the edge in a couple of objects, contrary to the
accretion disk predictions of Laor \etal, but qualitatively explicable
by effects found in the more detailed calculations of Blaes \& Agol
(1996);  see also Agol \& Blaes (1996).  The previous studies of
intermediate redshift quasars (Koratkar \ea 1995 and Impey \ea 1995),
included 4 objects, one of which shows only a marginal detection of
polarization, while the remaining three objects show significant
polarization ($>$ a few percent) shortward of the Lyman edge.
 A study of three high redshift objects observed from the ground failed
to show any polarization changes at the edge position; most had tight
limits on the  polarization longward of the edge, but noisy data
shortward of the Lyman edge (Antonucci \ea 1996).  PG 1630+377 is the
only object yet observed that can be studied in any detail (Koratkar
\ea 1995; Paper I).  In this object $P$ rises rapidly shortward of the
edge, reaching 20\% by 1600 {\AA} (650 {\AA} rest wavelength).  The
Ly$\alpha$ emission line also shows a high (7.3\%) polarization at the
same position angle. Antonucci \ea (1996) discuss polarization
observations and other constraints on disk models in some detail.

Based on the small number of QSOs observed in the UV in polarization,
at the time of paper I, we could say little about whether high
polarization shortward of the Lyman edge is common.  Hence, we have
significantly expanded the UV polarization database by observing six
bright, $z > 1$ QSOs both below and above the Lyman edge. In this paper
we discuss these new spectropolarimetric ultraviolet observations from
the {\it Hubble Space Telescope} Faint Object Spectrograph ({\it
HST/FOS}).  A difference with respect to our previous study, however,
is that only two of the new objects were suspected to have partial
edges in total flux at the systemic redshift.  The rest were simply
selected because they show significant flux at short wavelengths.

\section{Observations}

In their IUE archival search, Koratkar, Kinney \& Bohlin (1992)
identified six QSOs with partial Lyman edges consistent with edges from
simple thin accretion disks.  The possible observational detection of
accretion disk edges in these low red-shift quasars indicates that we
are more likely to see the accretion disk, and thus detect the
polarization changes, as predicted by theory. Also since these objects
are low redshift AGNs we expect the UV continuum shortward of
Ly$\alpha$ to be less affected by intervening absorption (the
Ly$\alpha$ forest).  Three objects of the IUE sample (PKS 0405-123, PG
1338+416, and PG 1630+377) have already been discussed in Koratkar \ea
(1995). Our present sample includes two more objects (PG 0117+213 and
PG 0743-673) from the IUE sample (see section 3 for the new evaluation
of these targets as possible partial Lyman edge candidates).  The
post-COSTAR FOS polarimetry capability does not extend to $\lambda <
$1600 \AA, hence the sixth object (PG 1538+447 at $z$ = 0.770) in the
IUE list was not observed.

The four other QSOs in the present sample are objects with significant
flux at the Lyman edge and shortward of the Lyman edge extending to
rest wavelengths $\leq$ 800 {\AA}.  In PG 1630+377 the rise in $P$
occurs at rest wavelengths of 770 {\AA} and below.  The shortest
wavelength which could be observed for the present sample was defined
either by (a) the shortest wavelength at which the FOS polarimeter
could work, which was 1600 {\AA}, or (b) the existence in most objects
of a sharp Lyman edge due to foreground gas at a lower redshift than
the QSO.  This latter constraint is often a significant one. The
present sample consists of six radio quiet QSOs which are given in
Table 1.

We observed the six objects, in Cycles 5 and 6, with the {\it HST/FOS}
spectropolarimeter to determine the amount of polarization on both
sides of the Lyman limit at 912 {\AA}.  The targets were acquired in
the 4.$''$3 aperture using the binary acquisition procedure of the
FOS.  All observations were obtained using the FOS blue detector and
the 1.0 arcsecond aperture. Details of the observations are given in
Table 1.  Except for PG 0117+213 all observations were obtained at 8
waveplate positions using the `B' waveplate, which is optimized for UV
observations. PG 0117+213 was observed at 4 waveplate positions, since
these observations were conducted before the change in the FOS
spectropolarimetric observing strategy recommended by the FOS team.
Because of the fewer waveplate positions, the PG 0117+213 data cannot
be corrected as accurately for the FOS instrumental polarization (see
details below).

The polarimetry calibrations are described in Allen \& Smith (1992).
The data were recalibrated using Allen's calibration program,``polar'',
because the current STSDAS pipeline calibration of polarization data is
inadequate for post-COSTAR data. The basic calibration procedure is
described in the {\it HST Data Handbook} (1997). The various
calibration steps are: (1) correction for dead diodes, (2) conversion
to count rates, (3) subtraction of the background, (4) flat field
correction, (5) computation of the wavelength solution, (6) conversion
to absolute flux, (7) correction for the wavelength-dependent FOS
instrumental polarization, and (8) calculation of the Stokes
parameters. In the post-calibration data reduction, the data were
binned in various ways, depending on the signal-to-noise ratio (S/N) of
the different data sets.

The introduction of the COSTAR mirrors introduced two $\sim$7\deg~
reflections in front of the FOS polarizer. These reflections therefore
convert some of the linear polarization into circular polarization
requiring a correction to the linear polarization data.  The residual
wavelength dependent instrumental polarization is $\buildrel < \over
\sim$ 0.2\% in observations of unpolarized standard stars after this
correction is performed.  Figure 1 shows the COSTAR induced
polarization in the unpolarized standard star BD+28\deg4211.  The
discontinuity seen in $U$ around 2700\AA\/ is due to the break between
the G190H and G270H gratings; each grating has a separate wavelength
dependent correction. Figure 2 shows the polarization spectrum of
BD+28\deg4211 after the COSTAR correction has been applied. The
instrumental polarization due to COSTAR can be corrected to
$\sim$0.08\%, except in the wavelength range 1800\AA\/ to 2100\AA\/
where the retardation of the waveplate goes through 180\deg (see Figure
2).  The other sources of error in polarization are due to photon
statistics of the observations and the error in the retardation
calibration (2\% of the linear polarization; Allen \& Smith 1992).  The
space craft ``jitter'', i.e., the telescope motion introduced due to
the thermal instability of the solar panels, is 0.007 arcsec (1$\sigma$
error).  This rarely introduced a photometric error of $>$ 1\% (because
the size of the FOS diodes was much larger than the jitter excursions).
Therefore, we find that space craft jitter does not produce
polarization greater than 1\%. Interstellar polarization is seen to
peak in the optical and then falls rapidly in the UV (Clayton \ea 1992;
Somerville \ea 1994), therefore we do not expect interstellar
polarization to be significant in our targets.

A note on how polarizations are calculated and manipulated should be
made here.  Since $P = \sqrt{Q^2 + U^2}$ is a positive-definite
quantity, it is often replaced by a ``debiased'' quantity, involving
not only $Q$ and $U$ but also their uncertainties. As Miller et al.
(1988) show, however, the error distribution of the debiasing method
normally used in optical polarimetry has a rather unsatisfactory form.
Simmons \& Stewart (1985) discuss various debiasing schemes. Throughout
this paper we choose to quote the standard, biased polarization, $P =
\sqrt{Q^2 + U^2}$.  This has the advantage of making the original
Stokes parameters more readily accessible to the reader.  We stress
that all {\it calculations}, such as correction for instrumental
polarizations or averaging over wavelength bins, are done on $Q$ and
$U$, which are essentially unbiased. The binned $Q$, $U$, $P$ and
$\theta$ are given in Table 2 and the Figures 3 -- 8 show the Stokes
parameters $Q/I$ (\%), $U/I$ (\%), and total flux for each object.  In
section 3 we have calculated the 95\% confidence level upper limits on
polarization using the confidence intervals derived by Simmons \&
Stewart (1985).

\section{Results}

The new objects generally have low polarization.  A brief description
of each one follows.

\subsection{PG 0117$+$213}

This object was fitted by a massive thin accretion disk by Laor
(1990).  It was selected in this sample because of the partial systemic
edge seen with IUE.  Our FOS observations show that the Lyman edge
region in this object is highly affected by intervening material at $z
\sim 1.36$ (noted by Koratkar, Kinney \& Bohlin 1992), which has
associated Ly$\alpha$ and Ly$\beta$ absorption lines (see Figure 3).

Longward of the Lyman edge there are hints of polarization.  Since the
position angle is approximately constant with wavelength, one could
combine all the bins longward of the Lyman edge and get polarization
$P$ = 0.8$\pm$0.5\% at $\theta$ = 116\deg$\pm$9\deg, but the
uncertainty here includes a large (0.4\%) contribution from
uncertainties in the post-COSTAR polarization correction, given that it
was taken in POLSCAN = 4 mode rather than POLSCAN = 8.  The bias
corrected optical polarization in PG 0117+213 is 0.44$\pm$0.22\% at
$\theta$ = 90\deg$\pm$15\deg~ (Berriman et al.1990).  The observed
optical and UV polarizations are essentially consistent.

No polarization is detected shortward of the Lyman edge. The upper
limits of the UV linear polarization (95\% confidence level as derived
by Simmons \& Stewart 1985), shortward and longward of the Lyman limit
are  $P$ = 1.4\% and 1.6\% respectively (see Table 3).

\subsection{0743$-$673}  

This object was also selected for an apparent partial systemic Lyman
edge in the IUE total-flux spectrum. However, our FOS data show no
Lyman edge.  Koratkar, Kinney \& Bohlin (1992) had noted that this
object was a ``weak'' candidate for a discontinuity at the Lyman edge
due to an accretion disk, because they could not evaluate the
significance of the Lyman edge discontinuity in the low signal-to-noise
IUE data.

No polarization is detected either shortward or longward of the Lyman
edge. (see Figure 4). The upper limits of the UV linear polarization
are $P$ = 7.5\% and 0.9\%, shortward and longward of the Lyman limit
respectively (see Table 3).

\subsection{PG 1247+267}  

Hints of polarization both shortward and longward of the edge position
are seen.  Since the position angles for all the bins are close, one
could combine them to produce $Q = -0.4\pm0.3$\%, $U = -0.5\pm0.3$\%
(see Figure 5).  This object is near the North Galactic Pole where
little interstellar polarization is expected.  PG 1247+267 shows
optical polarization of 0.41$\pm$0.18\% at $\theta$ = 97\deg$\pm$12\deg
~(Berriman \ea 1990) consistent with the {\it HST/FOS} UV data ($P$ =
0.6$\pm$0.3\% at $\theta$ = 118\deg$\pm$12\deg; upper limit on
polarization is 1.1\%; see Table 3).

\subsection{PG 1522+101}  

In this object, at longer wavelengths there is a marginal detection in
$Q$ at the very edge of the G270H grating at 3180 \AA\/ (see Figure 6
and Table 2).  The UV polarization in the wavelength range longward of
the Lyman edge is $P$ = 0.6$\pm$0.3\% at $\theta$ = 76\deg$\pm$15\deg.
If real, it is probably not from dust in the Galaxy at this sky
position and this far into the UV. The optical polarization from
Berriman \ea (1990) is 0.30$\pm$0.15\% at $\theta$ = 97\deg$\pm$14\deg,
which is consistent with the {\it HST/FOS} data.

At the shortest wavelength bin (rest wavelength of $\sim$714 \AA\/)
there may be a marginal detection of polarization. If real, this would
be at the wavelengths where we would have expected to see a rise in
polarization similar to that seen in Paper I.  Since the detection of
polarization is in a single wavelength bin and it has low significance,
the data are also consistent with no detection of polarization
shortward of the Lyman edge.

Once again, the upper limits of the UV linear polarization, shortward
and longward of the Lyman limit are  $P$ = 3.3\% and 1.4\%
respectively (see Table 3).

\subsection{PG 1718+481}

All data are consistent with zero polarization (see Figure 7), while
the optical polarization from Berriman \ea (1990) is 0.40$\pm$0.08\% at
$\theta$ = 76\deg$\pm$6\deg. The upper limits of the UV linear
polarization, shortward and longward of the Lyman limit are  $P$ =
1.0\% and 0.9\% respectively (see Table 3).

\subsection{UM 18 = 0002+051}  

All data are consistent with zero polarization (see Figure 8).  The UV
linear polarization upper limits, shortward and longward of the Lyman
limit are  $P$ = 2.8\% and 0.6\% respectively (see Table 3).

\section{Discussion and Conclusions}

Of the six QSOs identified by Koratkar, Kinney \& Bohlin (1992) as
candidate targets which have partial Lyman edges consistent with edges
from simple thin accretion disks, we now have spectropolarimetric
observations for five QSOs. We showed in section 3.2 that one of these
candidates, 0743$-$673 no longer qualifies as a partial Lyman edge
object.

There are only 13 high and intermediate redshift QSOs which have
spectropolarimetry observations shortward of the Lyman edge region.
These objects come from this paper, Koratkar \ea (1995), Impey \ea
(1995), and Antonucci \ea (1996).  At this point any detailed
statistical tests of polarization distributions are certainly not
warranted given the inhomogeneous selection criteria and data quality,
and the highly model-dependent predictions. Yet, Lyman edge
spectropolarimetry results can be summarized as follows:

\begin{itemize} 

\item{}Of the 13 objects only three objects (PG 1630+377, PG 1338+416
and PG 1222+228 from paper I and Impey \ea 1995) show significant
polarization at wavelengths shorter than 912\AA\/ (Lyman edge).  To
these three objects we can add one more marginal detection (PKS
0405$-$123 from paper I). All 13 objects in the sample show a
polarization signature which is inconsistent with any simple accretion
disk model. Furthermore, $\sim$30\% of the sample show a rise in
polarization shortward of the Lyman edge.  This observed rise in
polarization is qualitatively consistent with the disk models of Blaes
\& Agol (1996).  A number of different interpretations of the UV
signature have been given by Lee \& Blandford (1997), Shields, Wobus \&
Husfeld (1997) and by us in paper I.  We urge the interested reader to
consult those papers for more details.

\item{}There are a total of five objects (PG 0117+213 from the present
sample, PG1630+377, PG 1338+416, and PKS 0405$-$123 from paper I, and
0014+813 from Antonucci \ea 1996) which show candidate partial Lyman
absorption edges due to accretion disks at the systemic redshift in
total flux.  Of these five objects, two (PG 1630+377, PG 1338+416) have
sufficient signal-to-noise at rest wavelengths of $\leq$750\AA\/, and
show significant polarization shortward of the Lyman edge (at least a
few percent, detected at four sigma or greater significance). If the
rise in UV polarization seen in PG 1630+377 is characteristic of
objects with partial Lyman edges we need to observe rest wavelengths as
short as $\sim$700\AA.  The effective shortest rest wavelength observed
in PG 0117+213 and PKS 0405$-$123 is $\sim$800\AA\/. Thus in these
objects we could have missed the rise in polarization, although we do
have a marginal detection for PKS 0405$-$123. 0014+813 does not have
sufficient signal-to-noise shortward of the Lyman edge. To summarize,
of the objects that show candidate partial Lyman absorption edges in
total flux, $\sim$40\% show polarization shortward of the Lyman edge.

\item{}We have eight objects in the sample of 13 that do not show a
partial Lyman edge feature in total flux. Only one object out of these,
PG 1222+228, from Impey \ea (1995) shows significant polarization
shortward of the Lyman edge ($P$ = 4.6$\pm$0.9\%). The rest show no
detection of polarization in the 912\AA\/ spectral region. The linear
polarization upper limits shortward of the Lyman edge in the remaining
objects is \ltsima 4\%.

\item{}PG 1630+377 is the best studied object (see Paper I), and shows
UV polarization reaching 20\% at 650\AA\/ rest wavelength. Such high
degree of polarization is rare in (non-blazar) QSOs.

\end{itemize} 

Perhaps the simplest result of the current data is that it strengthens
the conclusion that quasars are generally not polarized significantly
just longward of the Lyman edge at $\sim$1000\AA\/.  This paper,
Koratkar \ea (1995), Impey \ea (1995), and Antonucci \ea (1996)
together present good observations of about 20 objects in the region
just longward of the Lyman edge, all with low UV polarization (\ltsima
1.5\%).  Further, there is no significant statistical wavelength
dependence to the polarization as predicted by the models of Laor et al
(1990). From these observations we conclude that simple plane-parallel
atmospheres with scattering-dominated opacity are not significant
sources of UV flux in quasars.

Recapitulating the previous discussions here briefly, we note that
models from the 1980s generally assumed that AGN accretion disks are
powered by viscous dissipation below the atmospheres, and that the
atmospheric opacities are dominated by electron scattering opacity in
the annuli that produce the rest optical and UV.  This results in 0\%
to 11.7\% polarization (Chandrasekhar 1960), depending on inclination,
and in a direction perpendicular to the symmetry axis of the disks.
Pioneering optical polarimetry observations showed much smaller
polarizations, which are {\it parallel} to the axes when the latter
could be inferred from a radio jet position angle (Stockman, Angel \&
Miley 1979; Antonucci 1988).  To explain the low observed polarization,
subsequent models by Laor et al (1990) suggested that electron
scattering was only important in the $\sim$1000-2000 \AA\/ range, with
the Lyman continuum and free-free absorption opacity dominating
shortwards and longwards of that interval respectively.  Other more
detailed calculations revealed that a lower fraction of absorption
opacity was required to reduce the predicted polarization than was
assumed by Laor et al; and that under rather special circumstances a
large polarization {\it parallel} to the disk axis could be produced
shortward of the Lyman edge (Blaes and Agol 1996, and references
therein).  The observed rise in UV polarization shortward of the Lyman
edge has been interpreted both in the context of accretion disk models
and non-disk related models. Here we do not further discuss the
polarization and depolarization mechanisms discussed in detail in Paper
I. An additional key complication in the accretion disk models, is that
AGN variability data require that the disk atmosphere is actually
illuminated from above, (e.g. Antonucci 1988, Sincell and Krolik 1996),
perhaps producing a strong polarization which cannot be calculated
rigorously without specification of the illumation geometry.

\acknowledgments

This work was supported through the {\it HST} GO grants GO-6109 and
GO-6705 provided by the Space Telescope Science Institute, which is
operated by the Association of Universities for research in Astronomy
Inc., under NASA contract NAS 5-26555. We thank the referee for his/her
comments and the thorough review of the paper.

\begin{table}
\begin{center}
\begin{tabular}{lcccc}
\multicolumn{5}{c}{{\bf Table 1:} Details of the Observations} \\
\tableline \tableline
Object & $z$ & Rest Wavelength & Grating & Exposure \\
       &     & range observed (\AA) &  Used   & Time (s) \\
\tableline
PG 0117+213 & 1.493 & 632 - 930  & G190H & 10500 \\
            &       & 892 - 1321 & G270H & 3100 \\
PG 0743-673 & 1.512 & 627 - 923  & G190H & 8693 \\
            &       & 885 - 1311 & G270H & 4420 \\
PG 1247+267 & 2.038 & 763 - 1084 & G270H & 3640 \\
PG 1522+101 & 1.321 & 679 - 999  & G190H & 6630 \\
            &       & 958 - 1419 & G270H & 1200 \\
PG 1718+481 & 1.084 & 755 - 1113 & G190H & 4660 \\
            &       & 1067 - 1580 & G270H & 1640 \\
UM 18       & 1.899 & 543 - 800  & G190H & 10950 \\
            &       & 767 -1136  & G270H & 3640 \\
\tableline
\end{tabular}
\end{center}
\end{table}
\clearpage

\begin{table}
\begin{center}
\begin{tabular}{lccccc}
\multicolumn{6}{c}{{\bf Table 2:} FOS Spectropolarimetry} \\
\tableline \tableline
Object & $\lambda$ (\AA) & Q (\%) & U (\%) & P (\%) & $\theta$ \\
\tableline
PG 0117+213 & & & & & \\
 & 2007$\pm$137 & 0.1$\pm$1.8  & -0.1$\pm$1.8 & 0.1$\pm$1.8 & ... \\
 &  2194$\pm$50 &  0.2$\pm$0.8 & -0.3$\pm$0.8 & 0.4$\pm$0.9 & ... \\
 &  2400$\pm$93 & -0.3$\pm$0.7 &  0.1$\pm$0.7 & 0.3$\pm$0.8 & ... \\
 &  2568$\pm$75 & -1.2$\pm$0.6 & -0.3$\pm$0.6 & 1.2$\pm$0.7 & 96\deg$\pm$14\deg \\
 & 2755$\pm$112 & -0.2$\pm$0.5 & -0.9$\pm$0.5 & 0.9$\pm$0.6 & ... \\
 & 3035$\pm$168 & -0.6$\pm$0.4 & -0.6$\pm$0.4 & 0.8$\pm$0.6 & 113\deg$\pm$14\deg \\
 & 3248$\pm$449 & -0.2$\pm$1.1 & -2.4$\pm$1.1 & 2.4$\pm$1.2 & 133\deg$\pm$13\deg \\
PG 0743-673 &  & & & & \\
 & 1727$\pm$157 & -1.8$\pm$2.5 & -0.6$\pm$2.5 & 1.8$\pm$2.5 & ... \\
 & 2072$\pm$188 & -1.6$\pm$0.9 &  0.5$\pm$0.9 & 1.7$\pm$0.9 &  82\deg$\pm$15\deg \\
 &  2417$\pm$94 & -0.1$\pm$0.8 & -0.2$\pm$0.8 & 0.2$\pm$0.8 & ... \\
 &  2587$\pm$75 & -0.6$\pm$0.8 & -0.1$\pm$0.8 & 0.6$\pm$0.8 & ... \\
 &  2744$\pm$83 &  0.4$\pm$0.8 & -0.8$\pm$0.8 & 0.9$\pm$0.8 & ... \\
 & 3059$\pm$234 &  0.2$\pm$0.4 & -0.5$\pm$0.4 & 0.5$\pm$0.4 & ... \\
PG 1247+267 & & & & & \\
 & 2363$\pm$143 & -0.7$\pm$0.6 & -1.4$\pm$0.6 & 1.6$\pm$0.6 & 121\deg$\pm$11\deg \\
 & 2620$\pm$114 &  0.4$\pm$0.6 & -0.2$\pm$0.6 & 0.4$\pm$0.6 &  ... \\
 & 2924$\pm$114 &  0.3$\pm$0.5 & -0.3$\pm$0.5 & 0.5$\pm$0.5 & ... \\
 & 3170$\pm$132 & -1.1$\pm$0.5 & -0.4$\pm$0.6 & 1.2$\pm$0.5 & 100\deg$\pm$14\deg \\
\tableline
\end{tabular}
\end{center}
\end{table}

\begin{table}
\begin{center}
\begin{tabular}{lccccc}
\multicolumn{6}{c}{{\bf Table 2 (continued):} FOS Spectropolarimetry} \\
\tableline \tableline
Object & $\lambda$ (\AA) & Q (\%) & U (\%) & P (\%) & $\theta$ \\
\tableline
PG 1522+101 & & & & & \\
 &  1658$\pm$82 &  3.8$\pm$1.9 & -2.8$\pm$1.9 & 4.7$\pm$1.9 & 162\deg$\pm$11\deg \\
 &  1828$\pm$87 &  0.6$\pm$1.1 & -0.1$\pm$1.1 & 0.6$\pm$1.1 & ... \\
 &  2005$\pm$87 & -0.1$\pm$0.8 & -0.6$\pm$0.8 & 0.6$\pm$0.8 & ... \\
 &  2233$\pm$85 & -0.4$\pm$0.5 & -1.0$\pm$0.5 & 1.1$\pm$0.5 & 123\deg$\pm$12\deg\\
 &  2272$\pm$48 & -0.3$\pm$1.4 &  0.1$\pm$1.4 & 0.3$\pm$1.4 & ... \\
 &  2408$\pm$87 & -0.5$\pm$0.8 &  1.0$\pm$0.8 & 1.2$\pm$0.8 & ... \\
 &  2582$\pm$87 & -0.2$\pm$0.8 &  0.9$\pm$0.8 & 0.9$\pm$0.8 & ... \\
 & 2866$\pm$197 & -0.2$\pm$0.5 & -0.3$\pm$0.5 & 0.3$\pm$0.5 & ... \\
 & 3179$\pm$115 & -2.0$\pm$0.9 &  0.1$\pm$0.9 & 2.0$\pm$0.9 & 89\deg$\pm$12\deg \\
PG 1718+481 & & & & & \\
 & 1726$\pm$150 & -0.2$\pm$0.7 &  0.3$\pm$0.7 & 0.3$\pm$0.7 & ...\\
 &  2006$\pm$78 &  0.2$\pm$0.5 & -0.6$\pm$0.5 & 0.6$\pm$0.5 & ...\\
 &  2162$\pm$78 &  0.3$\pm$0.4 &  0.8$\pm$0.4 & 0.8$\pm$0.4 & 34\deg$\pm$13\deg \\
 &  2280$\pm$40 &  0.9$\pm$0.5 &  0.3$\pm$0.5 & 0.9$\pm$0.5 & ...\\
 &  2284$\pm$60 &  1.9$\pm$0.7 & -0.5$\pm$0.7 & 2.0$\pm$0.7 & 172\deg$\pm$10\deg \\
 & 2542$\pm$198 &  0.2$\pm$0.3 &  0.3$\pm$0.3 & 0.4$\pm$0.3 & ...\\
 & 2918$\pm$177 &  0.5$\pm$0.3 &  0.3$\pm$0.4 & 0.5$\pm$0.3 & ...\\
 &  3194$\pm$99 &  0.1$\pm$0.6 &  0.5$\pm$0.6 & 0.5$\pm$0.6 & ...\\
\tableline
\end{tabular}
\end{center}
\end{table}

\begin{table}
\begin{center}
\begin{tabular}{lccccc}
\multicolumn{6}{c}{{\bf Table 2 (continued):} FOS Spectropolarimetry} \\
\tableline \tableline
Object & $\lambda$ (\AA) & Q (\%) & U (\%) & P (\%) & $\theta$ \\
\tableline
UM 18  & & & & & \\
 & 1858$\pm$157 & -0.9$\pm$2.0 & -1.2$\pm$2.0 & 1.5$\pm$2.0 & ...\\
 & 2167$\pm$152 &  0.4$\pm$0.8 & -0.8$\pm$0.8 & 0.9$\pm$0.8 & ...\\
 & 2416$\pm$193 &  1.5$\pm$0.8 & -0.6$\pm$0.8 & 1.6$\pm$0.8 & 169\deg$\pm$14\deg \\
 & 2790$\pm$109 &  0.2$\pm$0.8 &  0.2$\pm$0.8 & 0.3$\pm$0.8 & ...\\
 & 3008$\pm$109 & -0.2$\pm$0.8 &  0.5$\pm$0.8 & 0.5$\pm$0.8 & ...\\
 &  3205$\pm$88 &  0.0$\pm$1.0 & -1.3$\pm$1.0 & 1.3$\pm$1.0 & ... \\
\tableline
\tablecomments{
Values of $\theta$ with formal uncertainties larger than 15\deg 
~indicate less than 2$\sigma$ detections of the polarization and hence 
are omitted from the last column. The value of polarization 
$P$ = $\sqrt{Q^2+U^2}$ in this table. All values in the table have been
rounded off to the first decimal. The polarization uncertainty
for PG 0117+213 includes a large (0.4\%) contribution from uncertainties 
in the post-COSTAR polarization correction, given that it was taken in 
POLSCAN = 4 mode rather than POLSCAN = 8.}
\end{tabular}
\end{center}
\end{table}
\clearpage

\begin{table}
\begin{center}
\begin{tabular}{lccccc}
\multicolumn{6}{c}{{\bf Table 3:} Upper Limits on the UV Polarization } \\
\tableline \tableline
Object  & \multicolumn{2}{c}{\% P across 912\AA\/ } \\
 & Shortward & Longward \\
\tableline
PG 0117+213 & 1.4 & 1.6 \\
PG 0743-673 & 7.5 & 0.9 \\
PG 1247+267 & 1.1 & 1.1 \\
PG 1522+101 & 3.3 & 1.4 \\
PG 1718+481 & 1.0 & 0.9 \\
UM 18 & 2.8 & 0.6 \\
\tableline
\tablecomments{These upper limits of the UV linear polarization are 95\% confidence
levels as derived by Simmons \& Stewart 1985.}
\end{tabular}
\end{center}
\end{table}
\clearpage

\figcaption[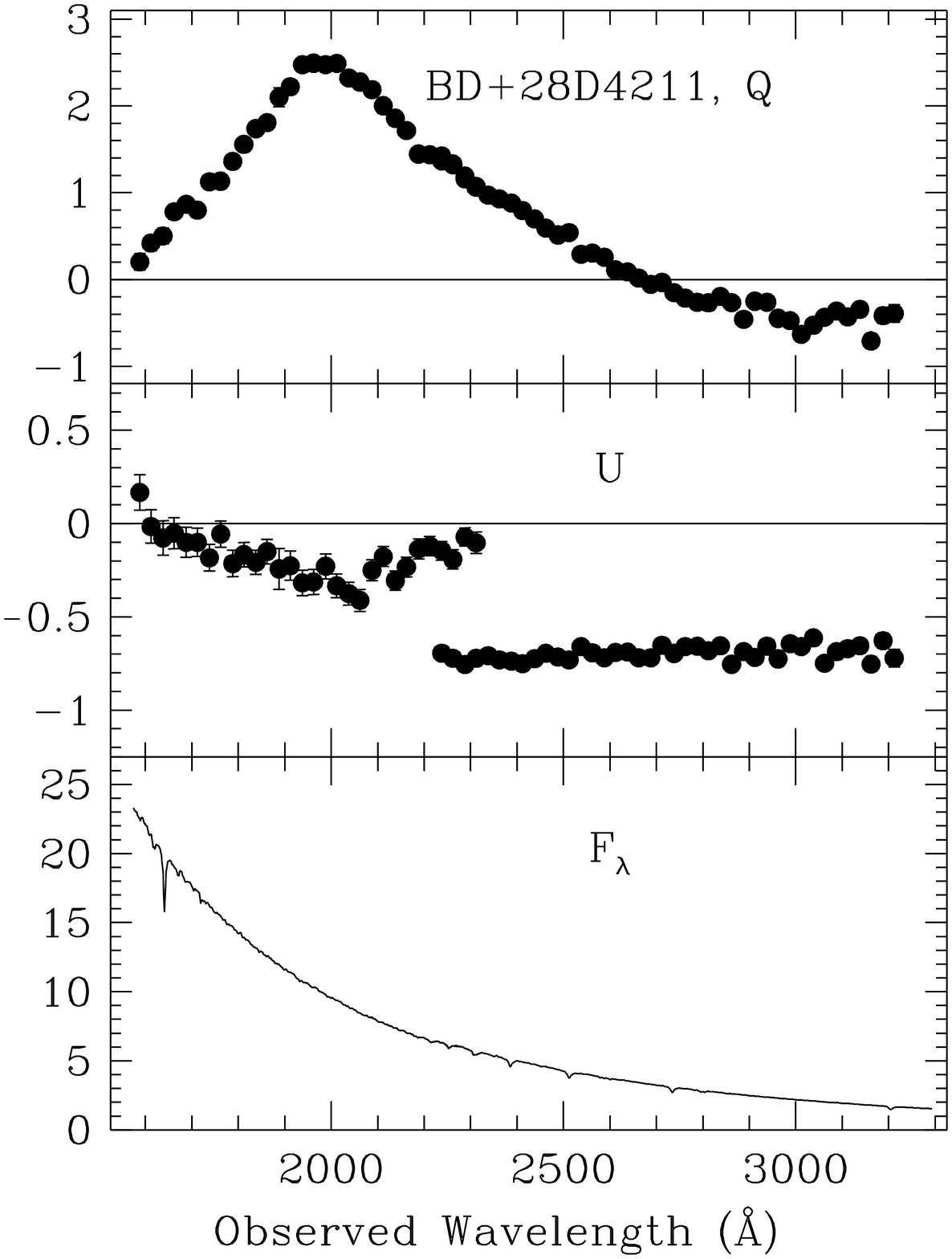]{FOS spectropolarimetry of the unpolarized standard
star BD+28\deg4211 uncorrected for post-COSTAR instrumental
polarization.  The normalized Stokes parameters $Q/I$ (\%) and $U/I$
(\%) are shown in the top two panels, binned in 25\AA\/ bins. The
bottom panel shows the total flux, F$_\lambda$, binned by 4 pixels =
1diode. F$_\lambda$ is given in units of 10$^{-12}$
erg~cm$^{-2}$~s$^{-1}$~\AA$^{-1}$. The post-COSTAR instrumental
polarization is clearly seen in $Q$ in the wavelength range 1800\AA\/
to 2100\AA\/. The discontinuity seen in $U$ around 2700\AA\/ is due to
the break between the G190H and G270H gratings.}

\figcaption[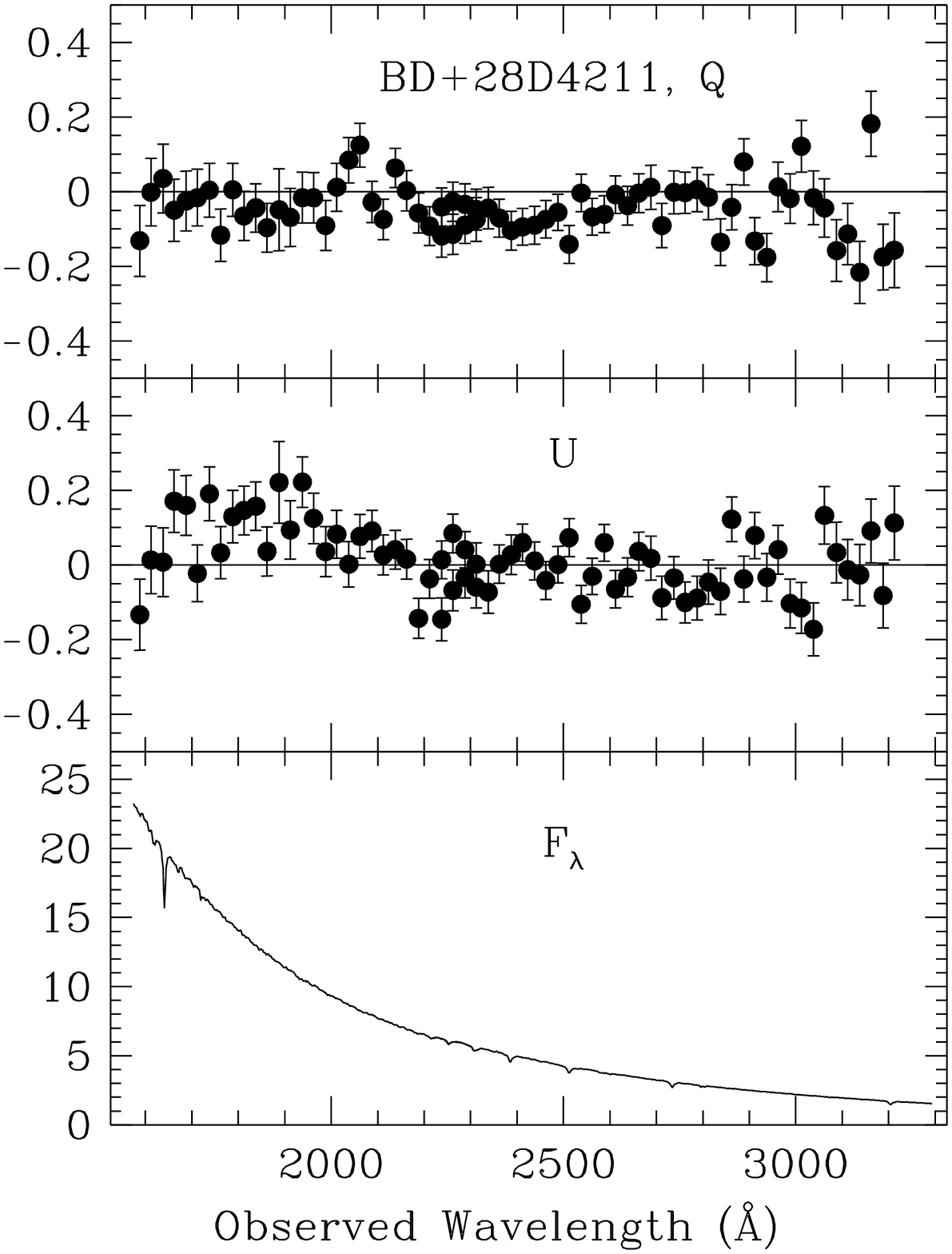]{Same as Figure 1, but corrected for post-COSTAR
instrumental polarization. The instrumental polarization due to COSTAR
can be corrected to $\sim$0.08\%.  }

\figcaption[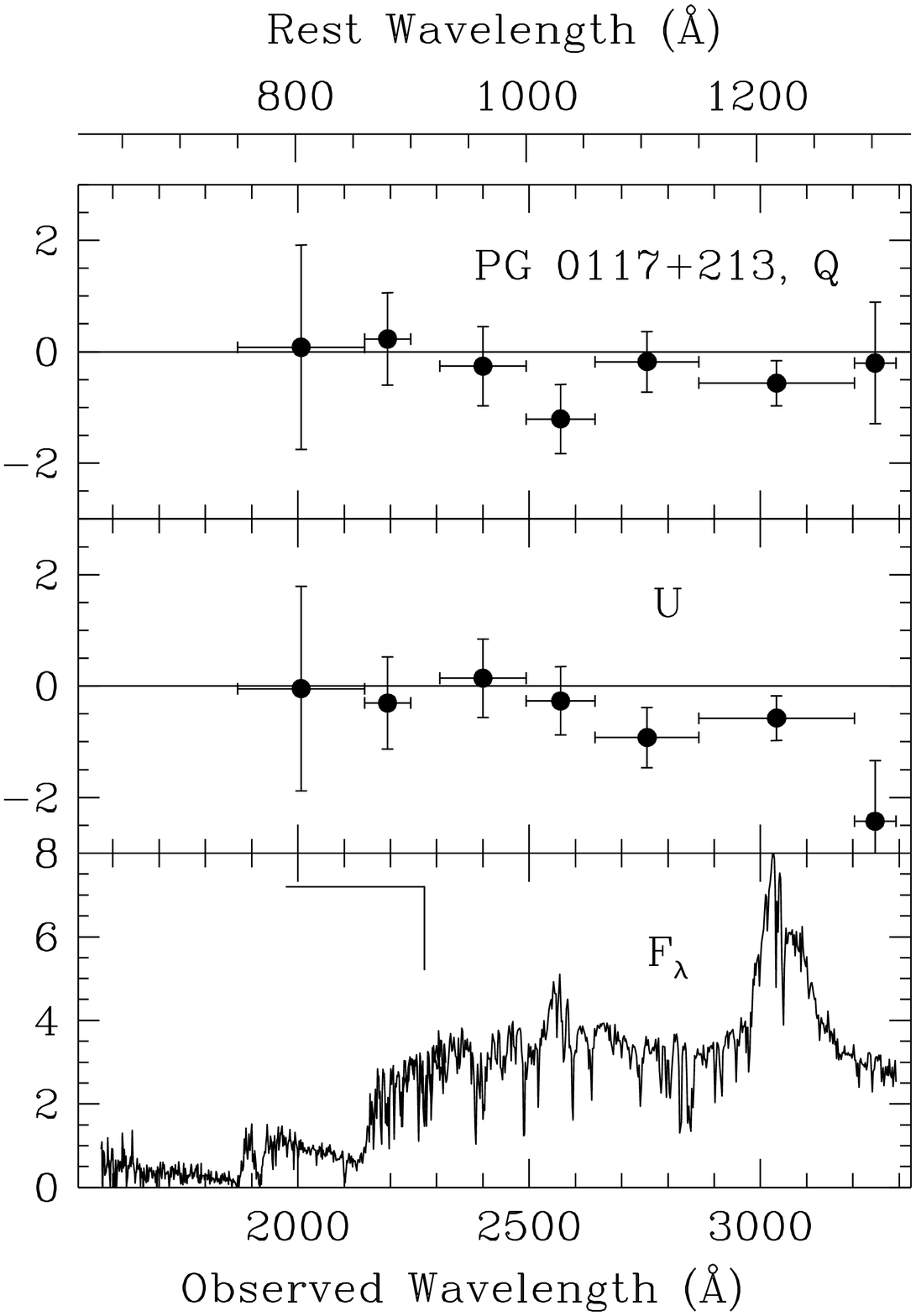]{FOS spectropolarimetry of PG 0117+213. The normalized
Stokes parameters $Q/I$ and $U/I$ are shown in the top two panels,
binned on either side of the Lyman edge represented by the right-angled
line. The bottom panel shows the total flux, F$_\lambda$, binned by 4
pixels = 1diode. F$_\lambda$ is given in units of 10$^{-15}$
erg~cm$^{-2}$~s$^{-1}$~\AA$^{-1}$.  }

\figcaption[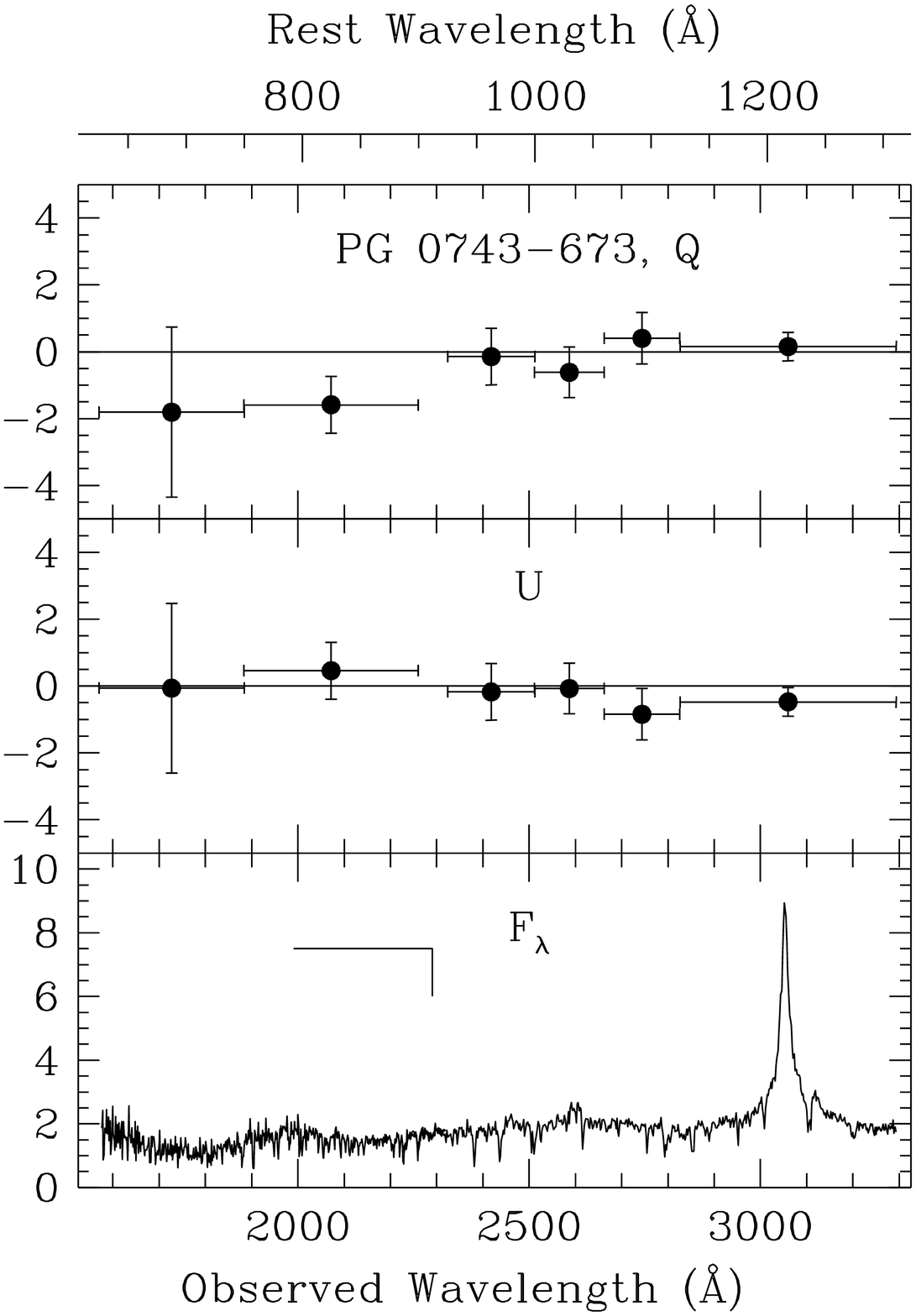]{Same as Figure 3 for 0743$-$673.}

\figcaption[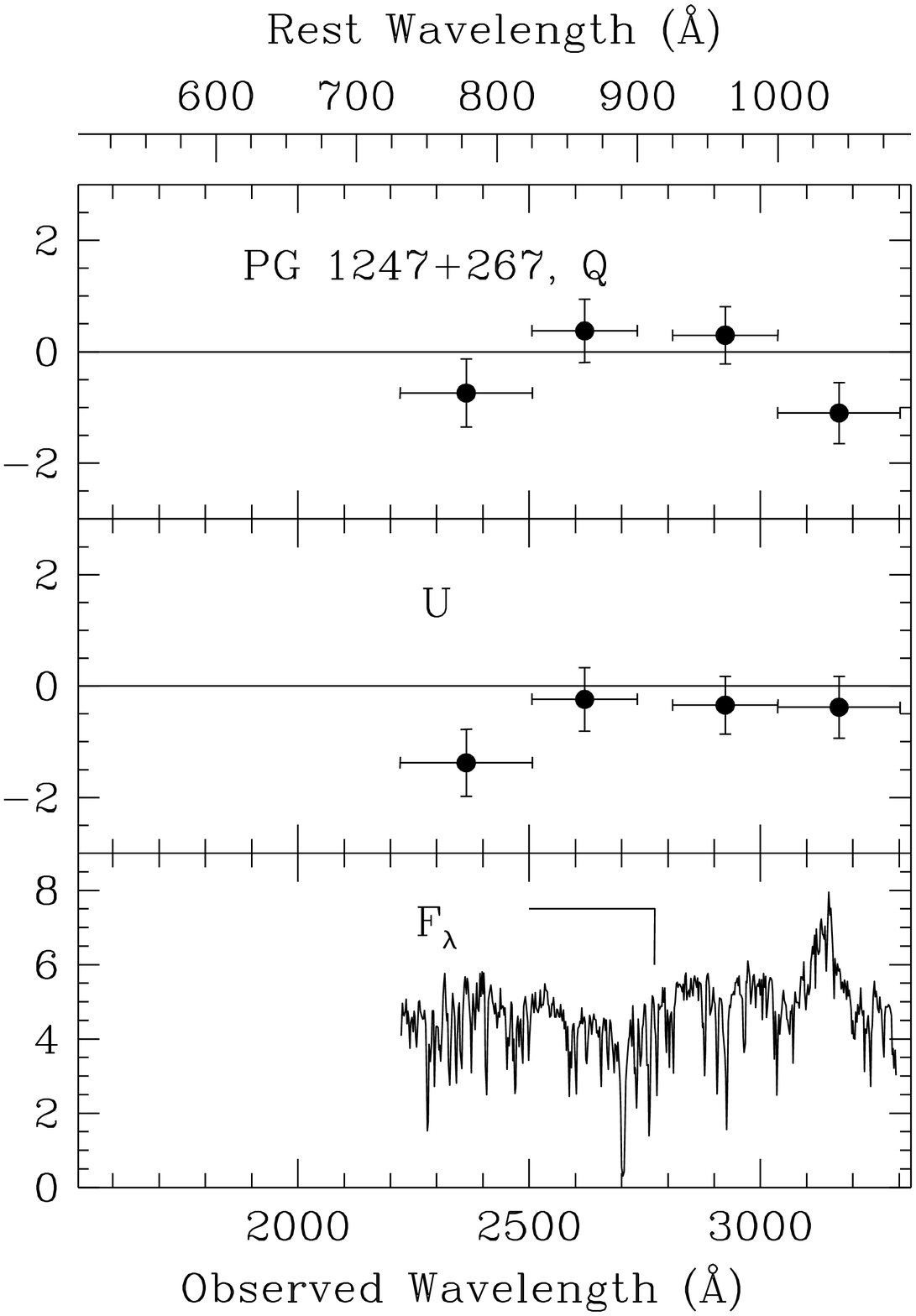]{Same as Figure 3 for PG 1247+267.}

\figcaption[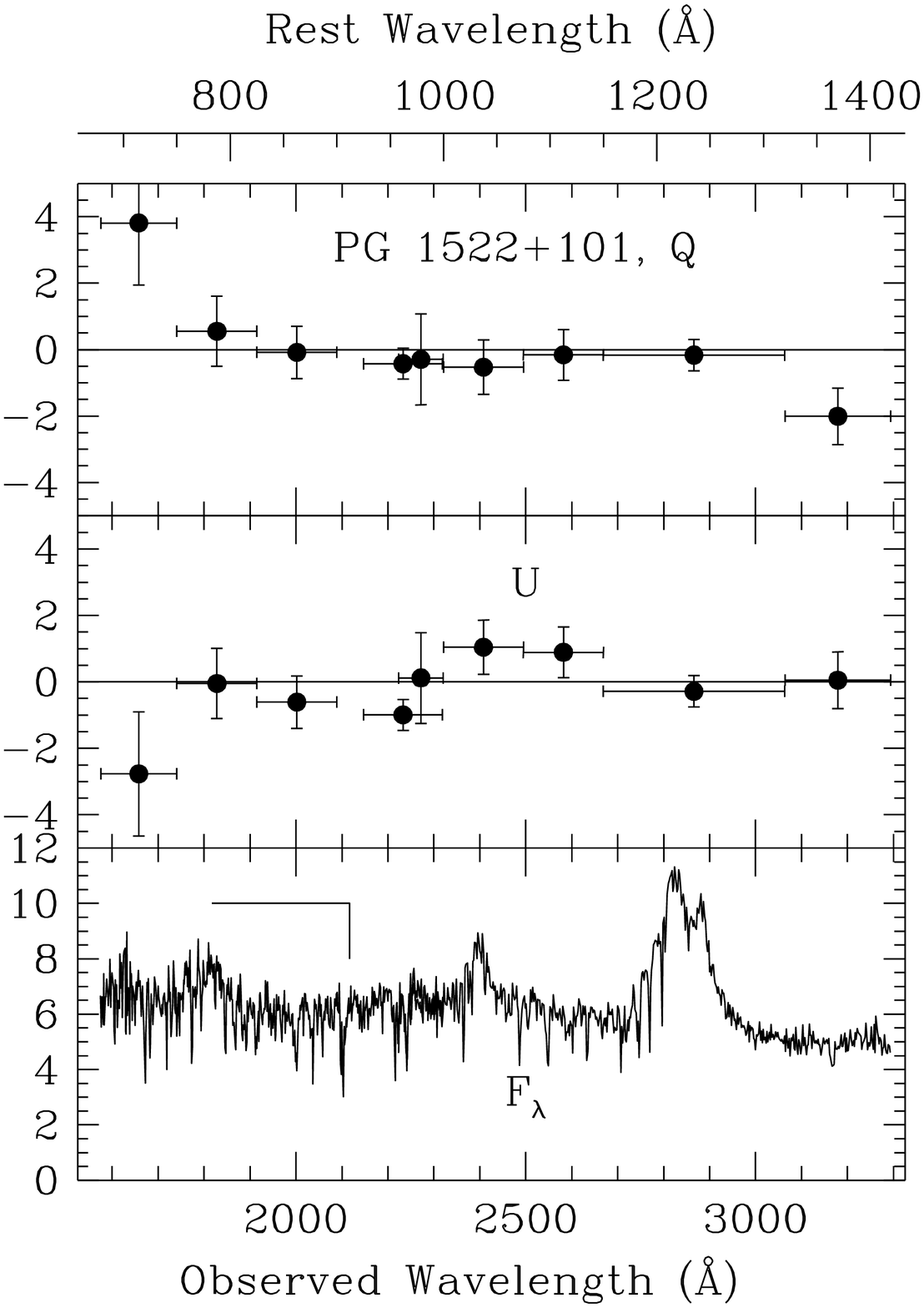]{Same as Figure 3 for PG 1522+101.}

\figcaption[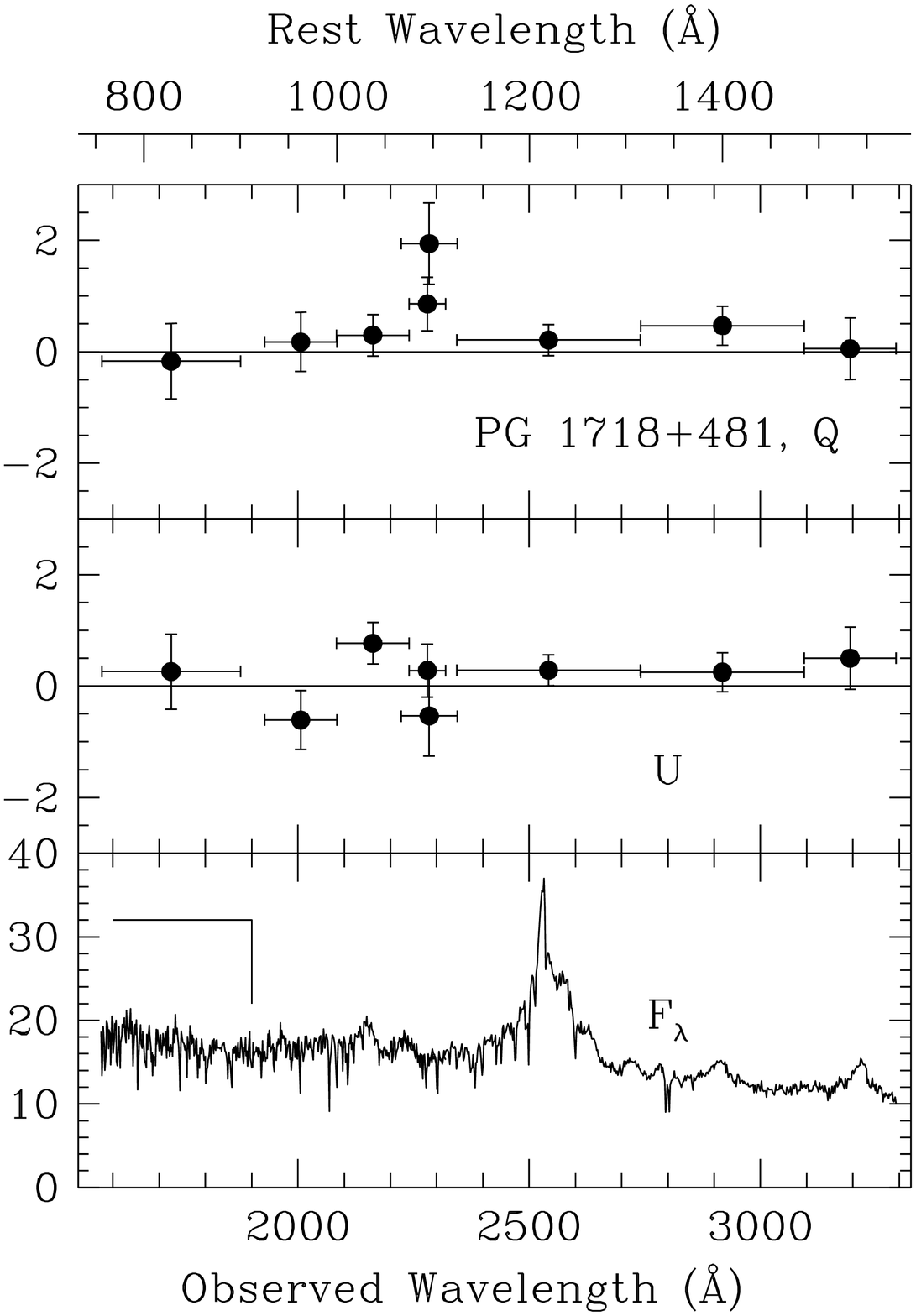]{Same as Figure 3 for PG 1718+481.}

\figcaption[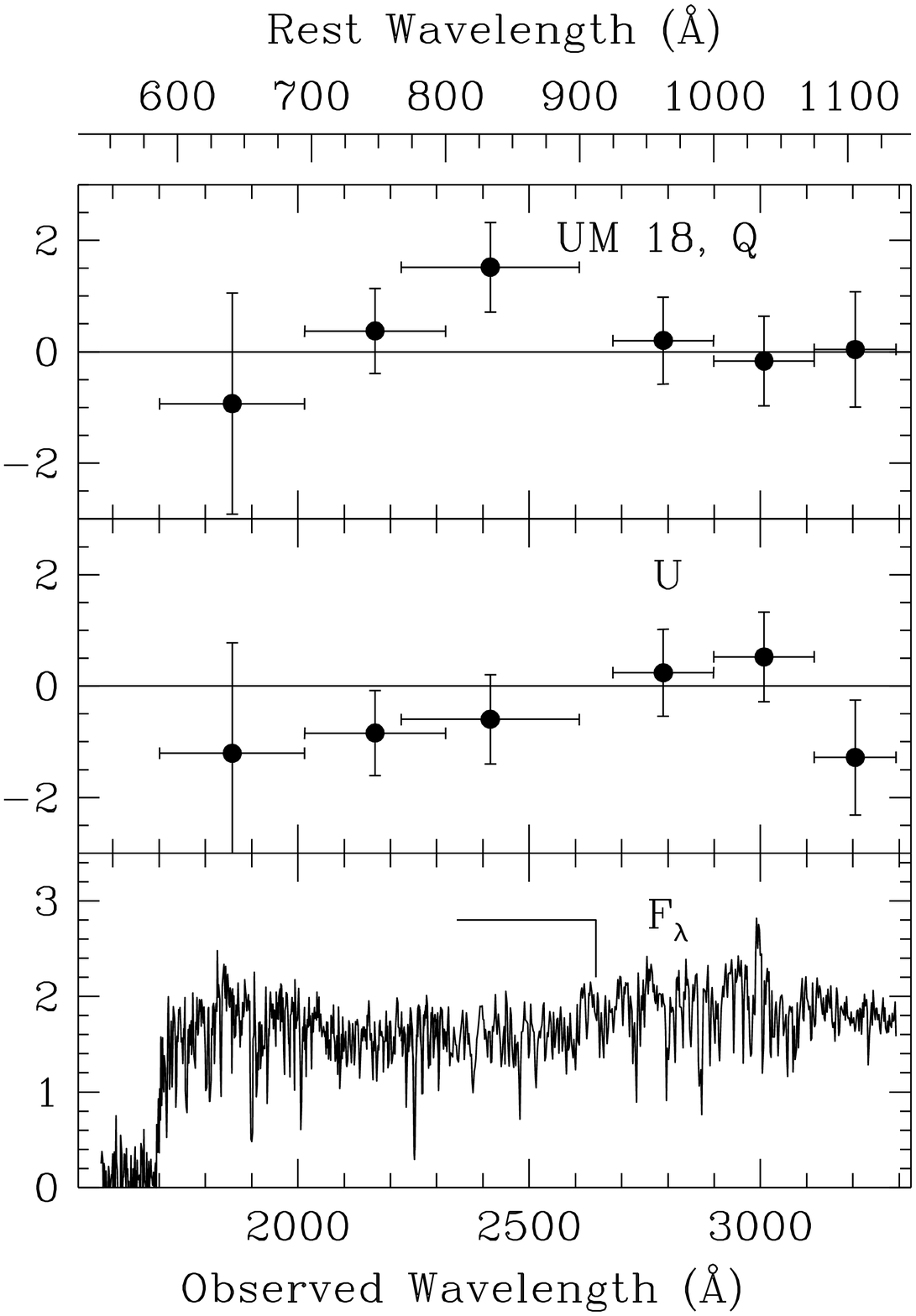]{Same as Figure 3 for UM18.}

\begin{figure}
\figurenum{1}
\plotone{fig1.ps}
\end{figure}
 
\begin{figure}
\figurenum{2}
\plotone{fig2.ps}
\end{figure}
 
\begin{figure}
\figurenum{3}
\plotone{fig3.ps}
\end{figure}
 
\begin{figure}
\figurenum{4}
\plotone{fig4.ps}
\end{figure}

\begin{figure}
\figurenum{5}
\plotone{fig5.ps}
\end{figure}
 
\begin{figure}
\figurenum{6}
\plotone{fig6.ps}
\end{figure}
 
\begin{figure}
\figurenum{7}
\plotone{fig7.ps}
\end{figure}
 
\begin{figure}
\figurenum{8}
\plotone{fig8.ps}
\end{figure}
 
\end{document}